\documentclass{emulateapj}
\usepackage{apjfonts,amsmath}

\newcommand{\bea}{\begin{eqnarray} }
\newcommand{\eea}{\end{eqnarray}}

\received{2004 June 21}
\accepted{2004 Sep }
\begin{document}

\title{Coevolution of the galactic cores and spiral galaxies}

\author{Takayuki R. SAITOH}
\affil{National Astronomical Observatory of Japan, Mitaka, Tokyo
181-8588, Japan\\
E-mail:saitoh.takayuki@nao.ac.jp}
\author{Keiichi WADA$^1$}
\affil{National Astronomical Observatory of Japan, Mitaka, Tokyo
181-8588, Japan\\
E-mail: wada.keiichi@nao.ac.jp}
\altaffiltext{1}{Department of Astronomical Science, The Graduate University for Advanced Studies, Osawa 2-21-1, Mitaka, Tokyo 181-8588, Japan.}

\begin{abstract}
Using high-resolution N-body/SPH simulations with $2\times 10^6$ particles, 
we investigate the evolution of stellar and gaseous galactic cores during the hierarchical 
formation of a spiral galaxy.
We find that the galactic core ($r < 300 $ pc) coevolves with the host galaxy. 
The average mass ratio between the baryonic core and the halo is nearly 
constant, $\sim $ 0.04 from $z \sim 10$ to $z \sim 2$. 
However, there are several `rapid-growing phases' during the evolution,
in which the rate of mass accretion to the central 
sub-kpc region is ten times higher ($\sim 1 M_\odot$ yr pc$^{-1}$) 
than the average accretion rate.  The rapid growth of the inner core 
is associated with the major merger events with a time-delay.
We also find that the spin-axis of the gas core frequently changes.
As a result, the angular momentum vector of the central part of the galaxy is independent of 
the rotation of the outer part. Our results suggest that if a constant fraction of
the baryonic mass in the central several 100 pc of a galaxy is converted into a
massive black hole, the black hole mass should correlate with the total mass of the 
galaxies.
\end{abstract}


\section{Introduction}

 The observed scaling relation between masses of black holes and galactic bulges 
 \citep{kor95, mer01, mcl02} strongly suggests that galaxies co-evolved with their galactic nuclei
during the hierarchical formation of galaxies. 
There has been a number of theoretical efforts to understand the origin of the relation between 
the black hole mass ($M_{\rm {BH}}$) and 
bulge mass or stellar velocity dispersion ($\sigma$) \citep[e.g.,][]{hae04}.
\citet{kau00} showed, using a semi-analytic approach, that many observational
aspects of galaxy evolution and quasars can be reproduced
if the supermassive black holes (SMBHs) are formed during major mergers of galaxies.
\citet{kaw02} proposed that the radiation drag arising from 
the radiation field in a galactic bulge could effectively work to accrete
gas to a seed back hole, and that it results in the correlation between the SMBHs and the bulges. 
\citet{gra04} presented a semi-analytic model for early coevolution of massive spheroidal
galaxies and AGNs, and claimed that feedback from supernovae and from AGNs determines
the scaling relation \citep[see also ][]{bur01}.
Introducing one free parameter, i.e. the returning fraction of 
AGN energy to the galactic gas, \citet{wyi03} showed that the ratio between the
BH mass and the stellar mass of the galactic spheroid should be $\sim 0.001$.
These studies rely on semi-analytic approaches with various assumptions and models.
On the other hand, direct numerical simulation of 
galaxy formation can be a complementary approach.
Using cosmological hydrodynamic simulations, Di Matteo et al.(2003) 
followed evolution of the gas, stars, and the dark matter in forming galaxies, and they 
found that the observed relation, $M_{\rm BH} \propto \sigma^4$, is reproduced
if the gas mass ($M_{\rm {gas}}$) in the bulges is linearly proportional to the black hole mass,
i.e. $M_{\rm BH} \sim 0.004 h^{-1} M_{\rm gas}$. 
This work directly shows gas mass fractions in dense regions of 
assembling galaxies, which can be used to reduce
the number of free parameters in semi-analytic approaches.
However, the numerical resolutions of Di Matteo et al.(2003) are quite limited;
The mass resolutions are $1.26\times 10^7 -1.43\times 10^8 h^{-1}M_\odot$, 
and the gravitational softening lengths are $4 - 9 h^{-1}$ kpc. 
With these resolutions, it is impossible to resolve
the structure of the central sub-kpc region of galaxies. One would expect that
the structure of the inner part of galaxies is more closely related with 
growth of the SMBHs. Therefore it is important to show a relation between
the baryonic mass in the central region of galaxies and assembling of the galaxies
in their formation history.

In this Letter, we show evolution of the central stellar and gaseous cores on a sub-kpc scale
during formation of a small (total mass is $10^{10} M_\odot $) spiral galaxy
through the hierarchical mergers, 
using high-resolution N-body/SPH simulations (one SPH particle has $10^3 M_\odot$ and
a softening length of $\sim 50$ pc).

%
%

\section{Methods and Models}\label{sec:Simulation}

The method used here is a standard numerical technique for 
simulating galaxy-formation.
Details of the simulations will appear elsewhere (Saitoh et al., in preparation).
We model formation of galaxies in the CDM universe 
using a standard hybrid N-body/hydrodynamic code
including the radiative cooling and star formation.
We use the Tree--algorithm with a special purpose computer, GRAPE (GRAvity PipE) \citep{sug90}, for
solving self-gravity, and SPH (Smoothed Particle Hydrodynamics) for gas dynamics.
We use a single GRAPE-5 board \citep{kaw00} connected with a host computer.
We employ a shear-free formulation of the artifical viscosity
to reduce unphysical angular momentum transfer in a shear flow \citep{bal95}.
The technique is also applied to simulations of galaxy formation \citep[e.g.][]{nav97}.
In order to model the multi-phase nature of the interstellar medium \citep[e.g.][]{wad01b},
we solve an energy equation with the radiative cooling below $10^4$ K and the inverse Comption cooling.
We set the gas has a primordial abundance with $X = 0.76$ and $Y = 0.24$
and we assume an ideal gas with $\gamma = 5/3$.
In the SPH simulations, the Jeans instability can be resolved correctly for 
masses larger than $2N_{\rm nb} m_{\rm SPH} $ \citep{bat97},
where $m_{\rm SPH}$ and $N_{\rm nb}$ are  mass of an SPH particle and 
the number of neighbor particles. Note that $m_{\rm SPH}$ does NOT simply represent
the mass resolution. 
We use $N_{\rm nb} = 50$ and $m_{\rm SPH} = 1.1 \times 10^3 M_{\odot}$,  therefore
we can resolve gravitational instability of a cloud mass $\ga 1.1\times 10^5 M_\odot$.
The temperature of this critical cloud is $\sim 500$ K for the number density $n \sim 100$ cm$^{-3}$.

The star formation algorithm we used is similar to one by \citet{kat92}.
If an SPH particle satisfies all the following conditions: 
(1) high density ($n_{\rm H} > 0.1 ~{\rm cm^{-3}}$),
(2) the regions are in virialized halos ($\rho_{\rm gas} > 200 \rho_{\rm BG}$), where $\rho_{\rm BG}$ is
the background density, (3) low temperature (T$<30,000$K), and (4) collapsing regions ($\nabla \cdot v < 0$),
then it is changed to a collisionless star particle, within $\sim$ 30 free-fall time
with the same velocity and mass of the SPH particle. 
In other word, a local star formation rate is assumed to be 
SFR $\equiv c_{*} m_{\rm {gas}}/\tau_{\rm {ff}}$  with 
$c_{*} = 1/30$ \citep[e.g.][]{aba03}.
One should note, however, that
these criteria for star formation in our simulations 
are almost equivalent to that for star formation 
in the Jeans unstable cloud.  
Dynamical and radiative feedback processes from star formation and
supernova explosions are not taken into account.

The number of dark matter and SPH particles is $N_{\rm DM} = N_{\rm SPH} = 1005600$.
The gravitational softening lengths are comovingly evolved
from the beginning of simulations to $z = 10$ \citep{gov97}.
After $z = 10$, the lengths are fixed at 52 pc for SPH and star particles
and 108 pc for DM particles.
Our mass and special resolutions are extremely fine compared with 
recent simulations of galaxy formation such as \citet{aba03} with
$\sim 10^6 M_\odot$ per a gas particle and a softening length of 0.5 kpc.

In order to use the number of particles as many as possible with our computer systems,
we employ a top-hat initial condition with an open boundary for a single galactic halo,
assuming the `Standard' CDM model.
$\Omega_0 = 1.0$, $\Omega_{\lambda} = 0.0$, 
$\Omega_{\rm b} = 0.1$, $h = H_0/{\rm km/s/Mpc} = 0.5$,
and $\sigma_8 = 0.63$.
The collapse epoch of the halo is set at $z_c \sim 3$.
The spin parameter of the halo is 0.05.
Since the total mass of the object in our simulation is small ($10^{10} M_\odot$) 
and the collapse epoch, $z_c \sim 3$, is earlier than that of much larger systems 
such as clusters of galaxies,
the results do not strongly depend on the employed cosmology,
especially with or without $\Lambda$ term.
The mass accretion phase is almost finished at $z \sim z_c$.
The employed boundary condition in our simulation would not be relevant 
to discuss evolution at low-$z$ (i.e. $z \ll z_c$).
In this letter, we discuss the evolution of galaxies until $z = 2$.
The initial condition is created by COSMICS \citep{ber01}.

%
\section{Results}
%


Figure 1 shows snapshots of gas and stars at redshift $z = 4.6$, 3.3,  and $z = 2.0$ showing
the assembling process of a small spiral galaxy. 
Two proto--galaxies are about to merge at $z=4.6$. 
At $z=3.3$, one of them is the small galaxy with a tidal tail seen at (+5 kpc,-3 kpc).
Many dwarf satellites will merge into a single spiral galaxy.
At $z=2$, spiral arms are formed in the gas disk, and spirals of young stars are also seen. 
These spiral structures are results of past major mergers like the one at $z=4.6$. 
Gaseous clumps are not prominent at this stage, but many stellar clusters are orbiting 
around the central galaxies. Tidally disrupted stellar clusters are also 
seen in this plot. 

\vspace{0.5cm}
\centerline{{\vbox{\epsfxsize=7.5cm\epsfbox{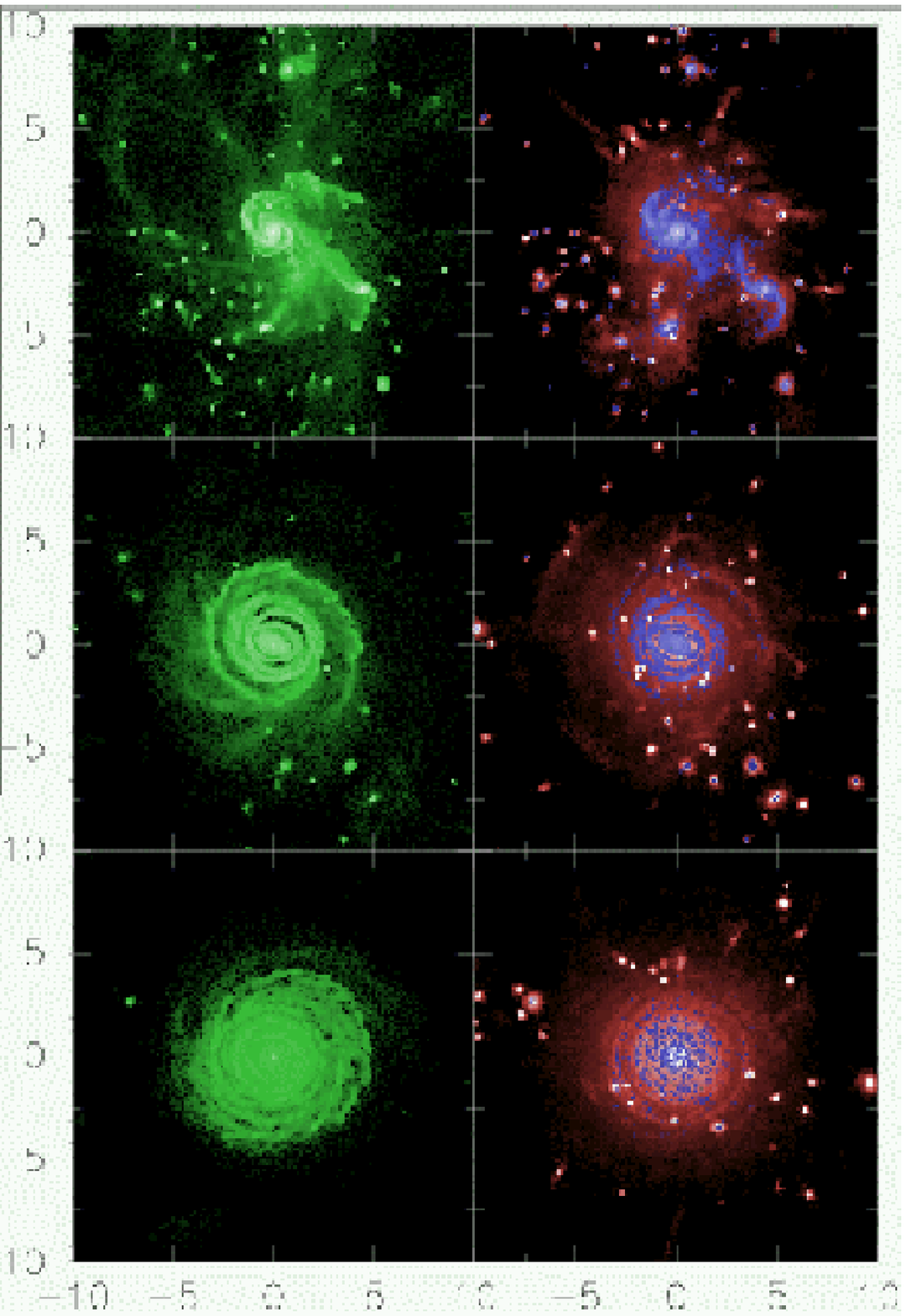}}}}
\figcaption{A snapshot of gas (left row) and stars (right row) in a forming galaxy 
at $z = 4.6$, $3.3$, and $2.0$. Stars older than 0.1 Gyr are represented as red.}
\vspace{0.5cm}

The three panels of Figure 2 are projections of a central 5 kpc of the gas disk 
from three different directions at $z = 2$.  
These clearly show that the inner part of the gas is not aligned with the outer one.  
We can also see that the gas is not distributed in a smoothed `disk', but rather in many rings.

\vspace{0.5cm}
\centerline{{\vbox{\epsfxsize=7.5cm\epsfbox{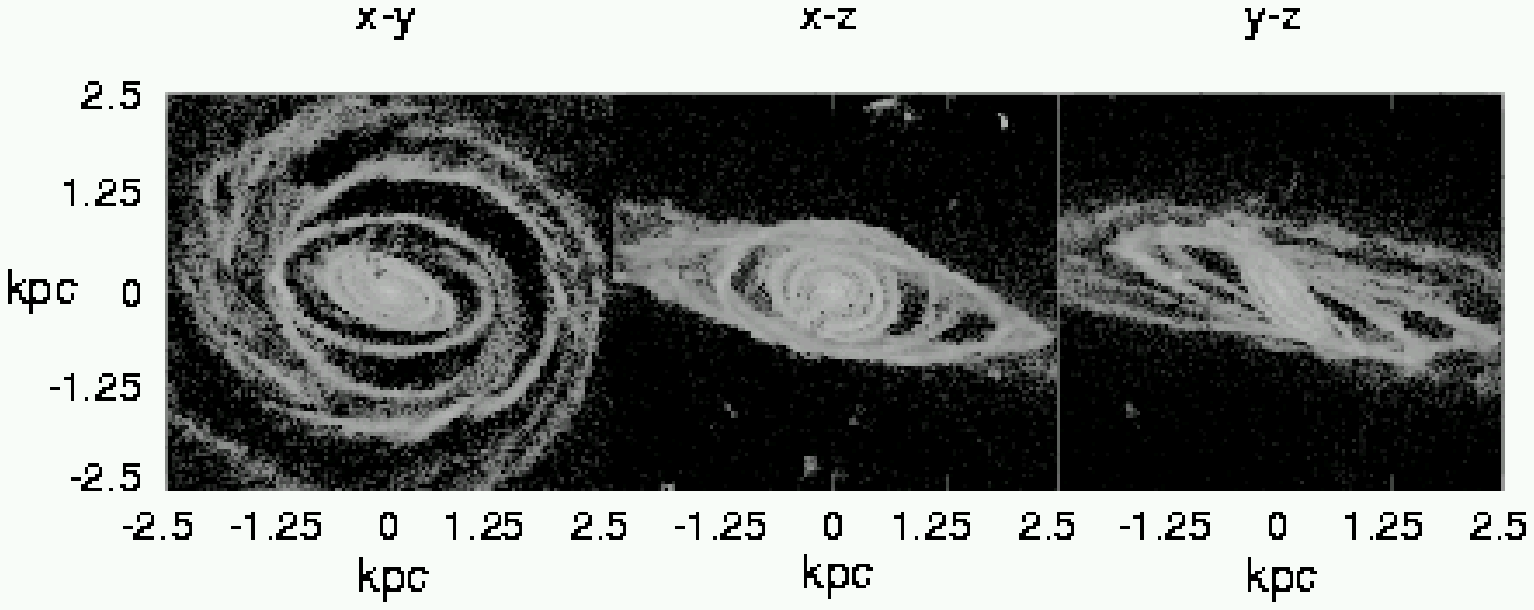}}}}
\figcaption{Three-dimensional projection of the gas inner 5 kpc at $z = 2.0$.}
\vspace{0.5cm}

We found that the baryonic core ($r < 0.3$ kpc) of the galaxy grows through
the mergings of satellites.   Two different processes of the growth are found: merging of 
gas-rich satellites and merging of stellar satellites.   The former process
dominates at high-z, and as a result of the merging events, a large amount of the gas
is supplied in the galactic central region.  This results in burst of star formation and
a stellar core is formed. On the other hand, at low-z, gases in the satellites are 
already consumed, therefore nearly pure-stellar satellites accrete into the host
galaxy, and they are eventually going to merge with the galactic core. 
More quantitatively, in our model, 34\% of the baryon in the core at $z = 2$
is originated in stars in satellites formed at $r >5$ kpc. 
49\% of  the baryon in the core at $z=2$ are formed form the accreted gas.
The angular momentum of the stellar core ($r < 0.3$ kpc)
at $z=2$ is about 7\% of the angular momentum of its gaseous/stellar progenitors
at $z =5$. For the gas phase, shocks and gravitational torque probably contribute
the transfer of the angular momentum, and stellar satellites are accreted due 
to dynamical friction with stars and the dark matter of the host galaxy.

Figure 3 is the evolution of the stellar and gaseous masses inside 300 pc from 
the center as a function of the halo mass (the mass inside a virial radius, which is
13 kpc at $z\sim 3$) of the spiral galaxy.
The baryonic core mass increases as the halo mass is developed due to mergers 
of smaller proto--galaxies. 
We find that the ratio between the baryon mass in the central region 
and the halo mass stays constant around 0.04 on average from
$z\sim 10$ to $z\sim 2$. One should note, however, that 
the process of mass increase is not smooth, but a step function-like. 
There are episodic `growing phases', 
where the accretion rate is about ten-times 
higher than the average value ($\sim 0.1 M_\odot$ yr$^{-1}$),
and each growing phase lasts for about $10^7$ yrs.
Each accretion phase seems to be triggered  by major mergers 
with some time delays, which are typically $10^7$ yrs.
The gases in the core are heat up at $T \sim 10^4$ K at most, 
then they are rapidly cooled below $10^4$ K.
Therefore the most accreted gases are cold and rapidly form stars in the core.
The drop of the dashed line for the gas at low-$z$ is 
owing to that the gas in the core region is consumed by the star formation.
Before $z \sim 3.5$, supply of gas due to mergers 
and the consumption of gas in the core are nearly balanced.
Accretion of the gas to the central region is decrease since $z \le 3.5$,
and it makes the core gas-poor.

\vspace{0.5cm}
\centerline{{\vbox{\epsfxsize=7.5cm\epsfbox{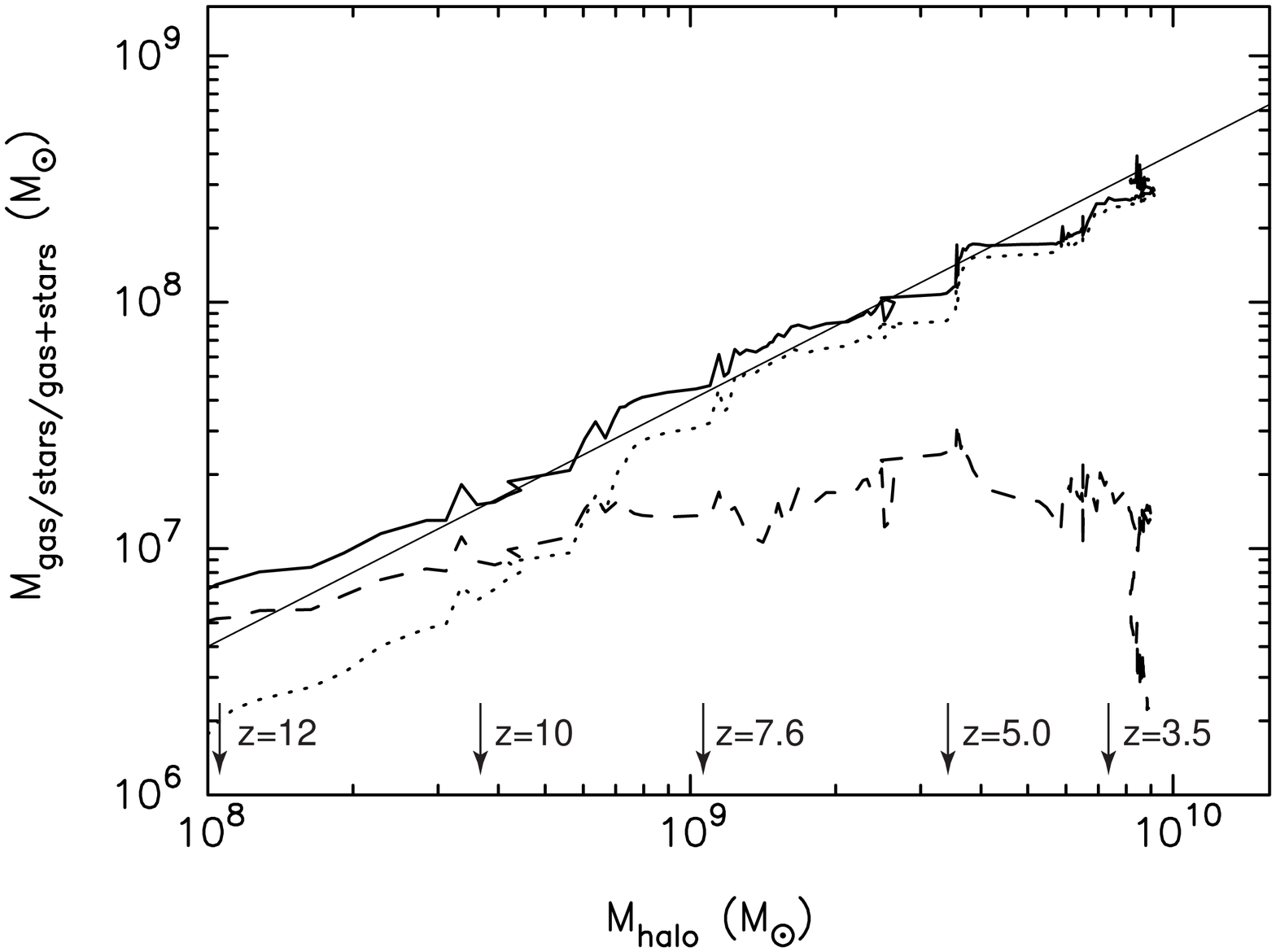}}}}
\figcaption{ Evolution of the gas (dashed line), stars (dotted line) 
and gas+stars (thick solid line) masses in the central 0.3 kpc 
against the halo mass ($r < r_{vir} = 13$ kpc at $z = 3$) from $z = 12$ to $z = 2$.
The thin solid line shows a constant mass ratio, i.e. $M_{\rm gas+stars}/M_{\rm halo} = 0.04$.
Arrows indicate the masses of the halo at laveled redshifts.}
\vspace{0.5cm}

Figure 4a and 4b are evolution of the normalized 
angular momentum vector of the gas core ($r < 0.3$ kpc and $r < 5$ kpc) 
projected on the $x-y$ plane. 
The two plots show that evolution of spin of the gas component depends on the radius.
The inner and outer parts rotate independently (see also Figure 2).
The spin vector of the gas 
in the core changes more extensively than does that of the outer part. 
In Figure 4a, at $z \sim 5$, the spin vector points in the $z$-direction, 
but it soon begins to fall, and at $z=4$, it points toward the $y$-axis, 
and then at $z=3.5$ it points toward the $x$-axis. 
The direction of the spin keeps changing towards $z=2.5$.
The sharp changes of the spins are because of frequent mergers 
followed by mass accretion to the center.
The spin vector of the gas supplied 
by each merger event is generally different from that of the already formed gas core,
and therefore the angular momentum of the gas core is changed extensively.

\vspace{0.5cm}
\centerline{{\vbox{\epsfxsize=7.5cm\epsfbox{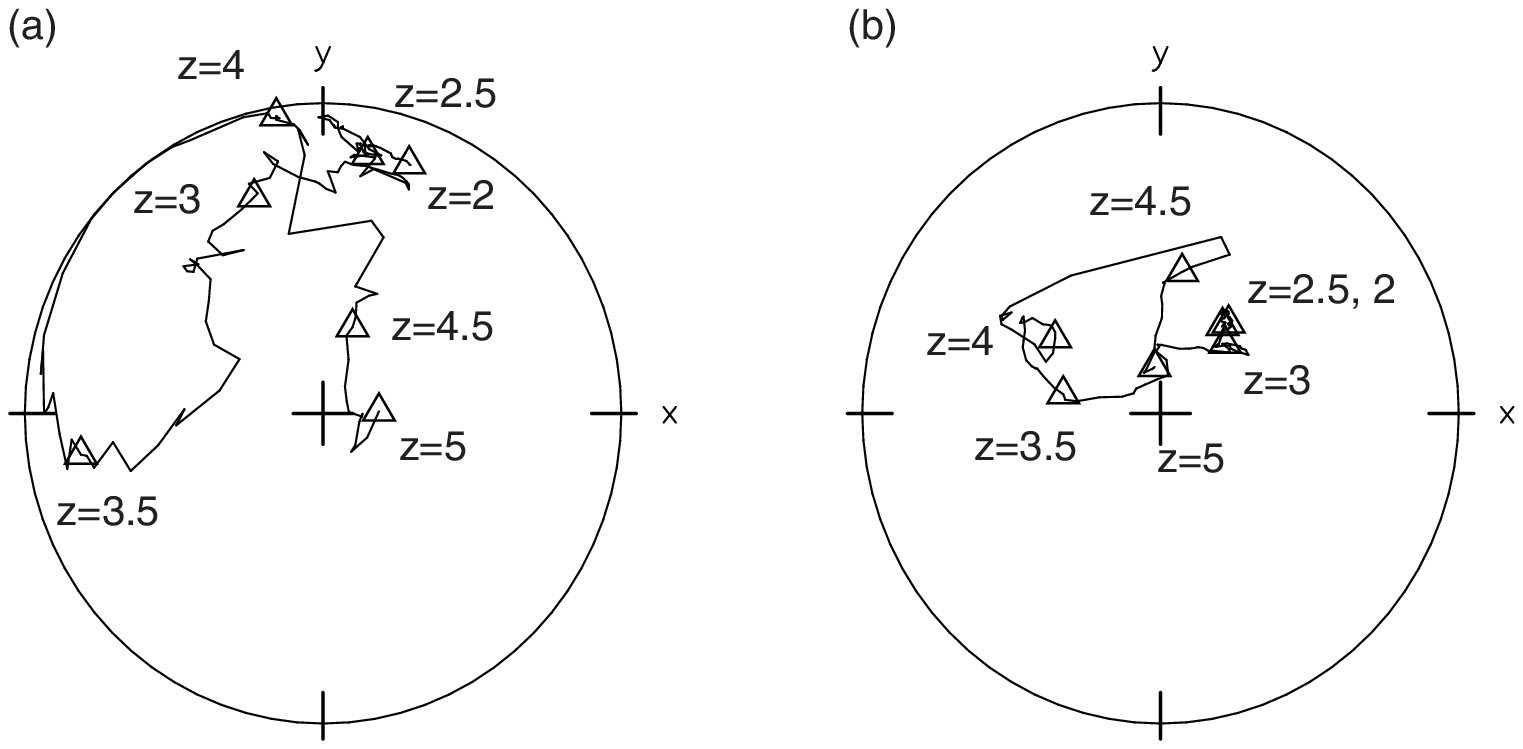}}}}
\figcaption{Trajectory of the spin vector of the gas component 
on an $x-y$ plane inside (a) 300 pc and (b) 5 kpc from the center. 
The cross at the center shows that the spin vector points
in the $z$-direction. If the trajectory follows the circle, 
it means that the spin vector is on the $x-y$ plane.}
\vspace{0.5cm}

%
\section{Conclusion and discussion}
%

Using high-resolution N-body/SPH simulation of formation of a spiral galaxy, 
we directly show that the galactic core ($r < 300$ pc) coevolves with
the galactic dark halo of a 10 kpc scale between $z\sim 10$ to $z\sim 2$.
The mass of the core increases with development of the halo mass, and, as a result,
the mass ratio between them stays nearly constant around 0.04.
There are episodic `growing phases', in which the rate of mass accretion to the
core is much higher than the average. Each growing phase lasts for $\sim 10^7$ yrs.
If the rapid increase of the gas mass in the central sub-kpc scale causes further mass
accretion to the nucleus, due to, for example, 
 an increase in turbulent viscosity \citep{wad02a} or radiation drag originating
in the nuclear starburst \citep{kaw02}, the growing phases might correspond to
quasar activities. If this is the case, the central SMBH also grows,
and as a result, the mass of the SMBH naturally correlates with the galactic 
total mass.

We also found that the angular momentum of the nuclear disk is decoupled with the large scale galactic spin.
This would be interesting in terms of the relation between the AGN jets and the rotational axes of disks
in Seyfert galaxies \citep{kin00} and dust disks in radio galaxies (Schmitt et al. 2000).
It is naturally expected that the spin axes of the cores are randomly directed if the
the gas core is formed by the accreted gas from past mergers.
If this is the case, 
the accretion to the SMBH and triggers of the AGNs should be
affected by the angular momentum of the accreted material.

Gas dynamics in two merging spiral galaxies have been studied using N-body/SPH simulations 
\citep[e.g.][]{bar91}. \citet{mih96} revealed that gas inflows are enhanced by major mergers,
and that the gaseous angular momentum is removed mainly by the gravitational torque.
Similarly, gaseous response in each merging process in our simulation 
is probably dominated by non-axisymmetric gravitational perturbations.
Minor mergers are also important to accrete gas into the central region \citep{mih94,tan96}.
In our simulations, we observe many minor mergers, and these also contribute to the
temporal increase in the accretion rate seen in Figure 3.

In order to investigate the effects of $c_{*}$, i.e. star formation rate (SFR),
for the evolution of the galactic core,
we run low-resolution tests, $N \sim 2 \times 10^5$ with $c_{*}$ = 0.33 and 0.033.
For the same $c_{*}(0.033)$, the evolution of the core 
between the simulation of $N \sim 2 \times 10^5$ is qualitatively similar 
to that in the high resolution model ($N \sim 2 \times 10^6$), as described in \$3.
Galactic cores evolve through major mergers, 
and then the masses of cores against the masses of halos keep almost constant during the evolution.
However, the growth of the core in the two test simulations are different 
for ten-times higher SFR, i.e. $c_{*} = 0.33$.
This high conversion rate allows gas to become stars more quickly
before gases form tightly bound core in each satellite by the effect of dissipation process.
As a result, forming stellar clusters are also bound more loosely and 
they are easily destroyed before they reach to the galactic core.
Most gas component is consumed by the high SFR, 
and therefore accretion of the gas, which is an important process 
to form the baryonic core at high-$z$, is not effective.
Thereby, baryon is less concentrate than the case of $c_{*} = 0.033$.

In the present model, we did not take into account effects of 
radiative and mechanical feedback from star forming activities.
Multiple supernova explosions followed 
by the galactic wind \citep[e.g.][]{mac99} could prevent accretion of the baryons toward
the galactic center \citep{bin01,spr03}.
In this sense, the mass fraction of the
baryonic core in our current model would be an "upper limit". 
Unfortunately, it is impossible to directly follow interaction 
between the multiple blastwaves and the ISM with
the current numerical resolution ($>$ 50 pc).  
\citet{wad02b} have performed three-dimensional, 
high resolution (0.25 pc) hydrodynamic 
simulations of multiple SNe in an inhomogeneous, turbulent medium in
the galactic central 100 pc region. They found that even if the supernova
rate is very high (0.8 SNe/yr), most of the dense gas stay around the
central region and form a massive `torus'. 
This is mainly because most of the gas mass is in a form of cold, dense gases,
and only a small fraction of the gas mass can escape from the region as a hot gas.
Therefore we expect that SNe would not affect significantly 
the gas mass in the central region of the dark halo potential.
On the other hand, stellar feedback would be more serious in the sub-clumps
surrounding the host galaxy. 
If this is the case, formation of the stellar cluster in the small dark halos is suppressed, 
and as a result the baryonic mass of the core would become smaller.  
Feedback from the back ground UV radiation would be 
also affect formation of small stellar systems \citep{sus04}.

\acknowledgments
The authors thank the anonymous refree for his/her fruitful coments and suggestions.
Numerical computations were carried out on GRAPE clusters (MUV) and Fujitsu VPP5000 at NAOJ.  
KW is supported by Grant-in-Aids for Scientific Research (no. 15684003) of JSPS.

\end{document}